\documentclass[12pt]{iopart}
\usepackage{iopams}  
\usepackage{graphicx}
\usepackage{comment}
\expandafter\let\csname equation*\endcsname\relax
\expandafter\let\csname endequation*\endcsname\relax

\usepackage{amsmath}
\usepackage{amssymb}
\usepackage{amsfonts}
\usepackage{url}
\usepackage[colorlinks=true,linkcolor=blue,citecolor=blue,urlcolor=blue,
hyperfootnotes]{hyperref}
\usepackage{bm}
\usepackage{array}
\usepackage{ragged2e}

\usepackage[font=small]{caption}

\usepackage[caption=false]{subfig} 

\newcommand{\be}{\begin{equation}}
\newcommand{\ee}{\end{equation}}

\newcommand{\ba}{\begin{eqnarray}}
\newcommand{\ea}{\end{eqnarray}}
\newcommand{\nn}{\nonumber}

\begin{document}

\title[Mass octupole and current quadrupole corrections from close hyperbolic encounters]{Mass octupole and current quadrupole corrections to gravitational wave emission from close hyperbolic encounters}

\author{Alexander Roskill$^{1,2}$, Marienza Caldarola$^{2}$, Sachiko Kuroyanagi$^{2,3}$, Savvas Nesseris$^{2}$}
\address{1.- Department of Physics, Imperial College London, Prince Consort Road, London SW7 2AZ, UK}
\address{2.- Instituto de F\'isica Te\'orica UAM-CSIC, Universidad Aut\'onoma de Madrid,
Cantoblanco, 28049 Madrid, Spain}
\address{3.- Department of Physics and Astrophysics, Nagoya University, Nagoya, 464-8602, Japan}
\ead{alex.roskill20@imperial.ac.uk\footnote{Corresponding author}, marienza.caldarola@csic.es, sachiko.kuroyanagi@csic.es, savvas.nesseris@csic.es}

\begin{abstract}
In this paper, we study the next-to-leading order corrections in the mass multipole expansion, i.e. the mass octupole and current quadrupole, to gravitational wave production by close hyperbolic encounters of compact objects. We find that the signal is again, as in the simple quadrupole case, a burst event with the majority of the released energy occurring during the closest approach. In particular, we investigate the relative contribution to the power, both in the time and frequency domains, and total energy emitted by each order in the mass multipole expansion in gravitational waves. To do so, we include in the quadrupole term its first order post-Newtonian correction, giving this a contribution to the power of the same order as that of the mass octupole and the current quadrupole. We find specific configurations of systems where these corrections could be important and should be taken into account when analysing burst events.
\end{abstract}
\begin{indented}
\item[]May 2024 \hfill IFT-UAM/CSIC-23-130
\end{indented}

\section{Introduction \label{sec:intro}}
The detection of gravitational waves (GWs) has become possible only in recent times, due to the extreme experimental sensitivity required. This was achieved by LIGO (Laser Interferometer Gravitational-Wave Observatory), with the initial recording of GW150914 in 2015~\cite{Abbott_2016}. Since the first observation, several subsequent recordings have been reported by the LIGO-Virgo-KAGRA collaboration~\cite{PhysRevLett.116.241103,PhysRevLett.118.221101, PhysRevLett.119.141101,PhysRevLett.119.161101,Abbott_2017_p, Abbott_2019_1,Abbott_2020,PhysRevD.102.043015,Abbott_2020_1,PhysRevLett.125.101102,Abbott_2021,LIGOScientific:2021qlt,theligoscientificcollaboration2021gwtc3}. Detection of GWs has consequently allowed for a plethora of new research in important areas of astrophysics and cosmology.

Despite the numerous detections of GWs that have been made to date, all of these observations have been for black holes (BHs) and neutron star (NS) merger events (or BH-NS~\cite{LIGOScientific:2021qlt}) because of the strength and duration of the signal, as well as its characteristic shape. The characterisation of signals different from those of binary systems is more difficult because they are too weak to be measured with current detectors, such as continuous GWs and stochastic GWs, or too short in duration to be clearly distinguishable from noise, as bursts. However, these sources of GWs are very likely to lead to new discoveries about the Universe. 
In particular, GW bursts have the potential to explore different phenomena of the Universe. Examples include core-collapse supernovae, neutron star excitation, non-linear memory effects, or cosmic string cusps and kinks.

In this paper, we consider one specific mechanism for generating GW bursts: close hyperbolic encounters (CHEs) of BHs~\cite{Turner:1977tm,Turner:1978ApJ,PhysRevD.5.1021}. These signals are expected to be strongly peaked, as GWs are produced when the objects accelerate past each other. The GW emission from a CHE event and possible detection by current detectors have been analysed, with no evidence of detection found~\cite{Morras:2021atg,Bini:2023gaj}. Despite the current difficulty in making good measurements of GWs produced by CHEs, it is believed that this could be possible with the new generation of GW detectors such as the Einstein Telescope~\cite{Maggiore_2020} and Cosmic Explorer~\cite{evans2021horizon}, due to their enhanced sensitivity~\cite{Mukherjee:2020hnm}. 

Studying the detection of CHE events and conducting a detailed analysis of their waveforms could provide opportunities for further tests of general relativity (GR). Current research into the relativistic two-body problem and GW modeling has provided an understanding of the dynamics and radiation emitted during these encounters~\cite{Bini_2012,PhysRevD.88.064051}. Nowadays, the dynamics of hyperbolic encounters can be described even in high-velocity/mass scenarios, incorporating non-linear effects, with high post-Newtonian (PN) and post-Minkowskian (PM) corrections~\cite{PhysRevD.96.064021,PhysRevD.104.084031,PhysRevD.107.024012,PhysRevD.107.064051,PhysRevD.108.124016}.
These events have also been studied using a plethora of novel approaches such as the effective-one-body formalism~\cite{Nagar:2020xsk,Andrade:2023trh,Nagar:2024dzj}, extending the analysis well beyond basic approximations. 

Moreover, significant progresses in numerical relativity simulations have also been made, especially to describe the merger regime~\cite{PhysRevD.89.081503,PhysRevD.107.124034}. Hyperbolic orbits correspond to either scattering processes or dynamic capture~\cite{East_2013,Gamba_2022}, and models to describe both scenarios have been proposed, some of which have been used in the data analysis of GW events.
Additionally, information about the CHE event rate is valuable for exploring the potential existence of primordial BHs~\cite{1966AZh....43..758Z,10.1093/mnras/152.1.75,1975ApJ...201....1C}, since these objects are likely to form dense clusters where CHEs are probable~\cite{Carr_2016,Clesse_2017,Trashorras:2020mwn} and the event rate may differ from that of astrophysical origin~\cite{Kocsis:2006hq,OLeary:2008myb,Mukherjee:2020hnm,Codazzo:2022aqj}. Observations of CHE events may also lead to the exploration of interesting phenomena, including spin induction~\cite{Nelson:2019czq,Jaraba:2021ces}, a stochastic GW background generation~\cite{Garcia-Bellido:2021jlq,Kerachian:2023gsa}, environmental effects~\cite{AbhishekChowdhuri:2023cle,Steane:2023gme}, etc. 

Therefore, given the significance of these phenomena, it is important to have accurate templates for GWs produced by CHEs. GW emission from hyperbolic encounters has been studied to first order using the mass quadrupole~\cite{CAPOZZIELLO_2008,De_Vittori_2012,Garc_a_Bellido_2017,Garc_a_Bellido_2018,DeVittori:2014psa,Cho_2018,Caldarola:2023ipo,Garc_a_Bellido_2017,Garc_a_Bellido_2018,Grobner:2020fnb}. Thus, in this paper, we specifically focus on furthering these current templates by analysing a binary system in a hyperbolic orbit and performing the next-to-leading order terms of the multipolar expansion, represented by the mass octupole and current quadrupole. We estimate their contribution to the GW strain and to the emitted power.

This paper is organised as follows. In Sec.~\ref{sec:II}, we describe the derivation of the GW strain and power in the context of the linearised theory, including the contribution of the lowest order term, i.e., the quadrupole radiation, and of the next-to-leading terms in the mass multipole expansion, i.e., the mass octupole and the current quadrupole radiation. In Sec.~\ref{sec:III}, we introduce the lowest-order correction in the PN expansion to the energy-momentum tensor and Newtonian equations of motion for hyperbolic encounters. This is necessary for the correct evaluation of the power. In Sec.~\ref{sec:IV}, we discuss the mass quadrupole contribution to GW waveform, while in Sec.~\ref{sec:V} the one of the mass octupole and current quadrupole for CHEs. In Sec.~\ref{sec:VI}, we provide the analysis for the different contribution to GW power and the total emitted energy, both in the time and frequency domains. We present our conclusion in Sec.~\ref{sec:Conclusions}.

\section{Theory of next-to-leading order corrections in the mass multipole expansion}
\label{sec:II}
To gain an intuition for calculating the GW strain and power to next-to-leading order (0.5PN $\equiv\mathcal{O}(v/c)$ corrections to the strain and 1PN $\equiv\mathcal{O}(v^2/c^2)$ corrections to the power), we begin by describing the calculations in the simple setting of the linearised theory and then set out how this gives the correct results in the full non-linear theory. In the linearised theory, we expand around the Minkowski metric as $g_{\mu\nu}=\eta_{\mu\nu}+h_{\mu\nu}$, where $h_{\mu\nu}\ll 1$ is a small perturbation associated with the GW. Using the Lorenz gauge $\partial^{\nu}\bar{h}_{\mu\nu}=0$, we find that 
\begin{equation}\label{eq:wave_equation}
 \Box ~\bar{h}_{\mu\nu}=-\frac{16\pi G}{c^4}T_{\mu\nu},
\end{equation}
where $\Box \equiv \partial^{\mu}\partial_{\mu}$ is the d'Alembertian operator and $\bar{h}_{\mu\nu}\equiv h_{\mu\nu}-\frac{1}{2}\eta_{\mu\nu}h$. We then find that the formal solution to Eq.~(\ref{eq:wave_equation}) is given by
\begin{equation}
 \bar{h}_{\mu\nu}(t,\textbf{x})=\frac{4G}{c^{4}}\int \mathrm{d}^{3}x'\frac{1}{|\textbf{x}-\textbf{x}'|}T_{\mu\nu}\left(t-\frac{|\textbf{x}-\textbf{x}'|}{c},\textbf{x}'\right),
\end{equation}
where the integral is evaluated over the source such that $|\textbf{x}'|\leq d_{s}$, where $d_{s}$ is the size of the source. Expanding the integrand for large distances to the point of measurement (far-field) and small velocities of the objects in the source~\cite{Maggiore_vol_1}, we find the multipole expansion
\begin{equation}\label{eq:multipole_exp}
 h^{\text{TT}}_{ij}=\frac{1}{D}\frac{4G}{c^{4}}\Lambda_{ij,kl}\left(S^{kl}+\frac{n_m}{c}\dot{S}^{kl,m}+\dots\right),
\end{equation}
where $D$ is the distance from the origin of the source to the point of measurement, $\Lambda$ is the projection operator which puts the metric into the transverse-traceless (TT) gauge, and
\begin{align}\label{eq:Skl}
 S^{ij}&=\int \mathrm{d}^{3}x~ T^{ij}, \\
 S^{ij,k}&=\int \mathrm{d}^{3}x~ T^{ij}x^{k}.
\end{align}
These integrals are evaluated at the retarded time ${t_{\text{ret}}=t-D/c}$ and the higher-order terms are defined similarly. Moreover, we have used the conservation of the energy-momentum tensor such that the physics is determined solely by the spatial-part of the metric (hence the relabeling of $h_{\mu\nu}$ to $h_{ij}$). The leading term in the multipole expansion of Eq.~(\ref{eq:multipole_exp}) is the mass quadrupole contribution, which is written as 
\begin{equation}\label{eq:h_quad}
 \left(h^{\text{TT}}_{ij}\right)_{\text{quad}}=\frac{1}{D}\frac{2G}{c^{4}}\Lambda_{ij,kl}\ddot{M}^{kl},
\end{equation}
where $\ddot{M}^{kl}=2S^{kl}$ is the second time derivative of the mass moment given by
\begin{equation}\label{eq:first_mass}
 M^{kl}=\frac{1}{c^2}\int \mathrm{d}^{3}x \,T^{00}x^{k}x^{l}.
\end{equation}
The mass quadrupole contribution is typically written in terms of the quadrupole moment 
\begin{equation}\label{eq:reduced_quadrupole}
 Q^{kl}\equiv M^{kl}-\frac{1}{3}\delta^{kl}M^{ii},
\end{equation}
resulting in the re-definition of Eq.~(\ref{eq:h_quad}) as 
\begin{equation}\label{eq:new_h_quad}
 \left(h^{\text{TT}}_{ij}\right)_{\text{quad}}=\frac{1}{D}\frac{2G}{c^{4}}\Lambda_{ij,kl}\ddot{Q}^{kl}.
\end{equation}
The next-to-leading order term of the multipole expansion of Eq.~(\ref{eq:multipole_exp}) is decomposed, as irreducible representations of $SO(3)$, into a mass octupole and current quadrupole contribution as follows 
\begin{equation}\label{eq:next_to_leading}
\begin{split}
 \left(h^{\text{TT}}_{ij}\right)_{\text{next-to-leading}}&=\frac{1}{D}\frac{4G}{c^{5}}\Lambda_{ij,kl}n_{m}\dot{S}^{kl,m}\\
 &= \frac{1}{D}\frac{2G}{3c^{5}}\Lambda_{ij,kl}n_{m}\dddot{M}^{klm} +\frac{1}{D}\frac{4G}{3c^{5}}\Lambda_{ij,kl}n_{m}\ddot{\mathcal{Z}}^{klm},
\end{split}
\end{equation}
where the first and second terms in the final equality correspond to the mass octupole and current quadrupole contributions, respectively. The quantity $\dddot{M}^{klm}$ is the third time derivative of the mass moment
\begin{equation}\label{eq:second_mass}
 M^{klm}=\frac{1}{c^2}\int \mathrm{d}^{3}x \,T^{00}x^{k}x^{l}x^{m}.
\end{equation}
Furthermore, $\ddot{\mathcal{Z}}^{klm}$ is the second time derivative of the current quadrupole moment
\begin{equation}
\mathcal{Z}^{klm}\equiv{P}^{k,lm}+{P}^{l,km}-2{P}^{m,kl},
\end{equation}
where 
\begin{equation}\label{eq:first_current}
 P^{i,jk} = \frac{1}{c}\int \mathrm{d}^{3}x \ T^{0i}x^{j}x^{k}.
\end{equation}
The mass octupole contribution is often re-written similarly to Eq.~(\ref{eq:new_h_quad}) in terms of the mass octupole moment 
\begin{equation}\label{eq:mass_octupole}
 \begin{split}
 \mathcal{O}^{klm}\equiv&M^{klm}-\frac{1}{5}\biggl[\delta^{kl}M^{k'k'm}
 +\delta^{km}M^{k'lk'}+\delta^{lm}M^{kk'k'}\biggl],
 \end{split}
\end{equation}
such that the two parts of the next-to-leading order contribution to the perturbation to the metric associated with the GW are
\begin{align}
 &\label{eq:mass_octupole_contribution}\left(h_{ij}^{\text{TT}}\right)_{\text{oct}}=\frac{1}{D}\,\frac{2G}{3c^5}\,\Lambda_{ij,kl}\,n_{m}\,\dddot{\mathcal{O}}^{klm},\\
 &\label{eq:current_quad_contribution}\left(h_{ij}^{\text{TT}}\right)_{\text{cur.quad.}}=\frac{1}{D}\,\frac{4G}{3c^5}\,\Lambda_{ij,kl}\,n_{m}\,\ddot{\mathcal{Z}}^{klm}.
\end{align}
The power carried by a GW is given by 
\begin{equation}
 P = \frac{c^3 D^2}{32\pi G}\int \mathrm{d}\Omega~\langle \dot{h}_{ij}\dot{h}^{ij}\rangle,
\end{equation}
which upon integrating over the solid angle and using Eqs.~(\ref{eq:new_h_quad}), (\ref{eq:mass_octupole_contribution}) and (\ref{eq:current_quad_contribution}), we find that the power up to next-to-leading order is 
\begin{equation}\label{eq:power_nextto}
 \begin{split}
 P &= \frac{G}{c^5}\biggl[\frac{1}{5}\langle\dddot{Q}_{ij}\dddot{Q}^{ij}\rangle+\frac{16}{45}\frac{1}{c^2}\langle\dddot{\mathcal{J}}_{ij}\dddot{\mathcal{J}}^{ij}\rangle
 +\frac{1}{189}\frac{1}{c^2}\langle\ddddot{\mathcal{O}}_{ijk}\ddddot{\mathcal{O}}^{ijk}\rangle+\mathcal{O}\left(\frac{v^4}{c^4}\right)\biggl],
 \end{split}
\end{equation}
where ${\mathcal{J}^{ij}\equiv \frac{1}{2}(J^{i,j}+J^{j,i})}$ is a symmetric traceless tensor constructed from 
\begin{equation}
 J^{i,j} = \frac{1}{c}\int\mathrm{d}^{3}x\ x^{j}\epsilon^{ipk}\,x^{p}\,T^{0k}.
\end{equation}
It is clear that the energy-momentum tensors and coordinates used in the definitions of the mass and current moments, Eqs.~(\ref{eq:first_mass}),(\ref{eq:second_mass}) and (\ref{eq:first_current}), must be valid to the order which the GW strain and power are accurate to. Neglecting the effects of back-scattering of GW on the masses in orbit (these occur at 2.5PN so are not relevant here), we find the general form of the PN corrections to $T^{00}$ and $T^{0i}$ to look like
\begin{align}
T^{00} &= {}^{(0)}T^{00}+ {}^{(2)}T^{00} + \mathcal{O}\left(v^4/c^4\right),\label{eq:T00_exp} \\
T^{0i} &= {}^{(1)}T^{0i}+ {}^{(3)}T^{0i} +\mathcal{O}\left(v^5/c^5\right),\label{eq:T0i_exp}
\end{align}
where the $(n)$ indices denote the expansion order in $v/c$. Calculations for the strain can be achieved at the 0.5PN order without PN corrections to the energy-momentum or trajectory since the corrections first appear  at the 1PN level.

However, hereafter, we include the term ${}^{(2)}T^{00}$ in Eq.~(\ref{eq:T00_exp}) for the calculations of the quadrupole power in Eq.~(\ref{eq:power_nextto}) since this appears at the same order as the mass octupole and current quadrupole contributions to the power. Additionally, the trajectory of the masses must also be to the correct PN order for the calculation of the power in Eq.~(\ref{eq:power_nextto}). We also require Eqs.~(\ref{eq:T00_exp}) and (\ref{eq:T0i_exp}) to find the trajectory to 1PN order~\cite{Cho_2018} (we solve for the PN metric and use it to construct a Lagrangian to extract the equations of motion) so that both the GW strain and power are consistent to the next-to-leading order. The details of how we achieve these corrections to the energy-momentum and orbit are described in Sec.~\ref{sec:III}.

It is worth noting that the formalism presented here can be retrieved at the next-to-leading order through the full formalism developed by Blanchet and Damour among others, where the relaxed Einstein equations are used~\cite{blanchet2024postnewtonian,Maggiore_vol_1}. At higher PN orders, the separate expansions of Eqs.~(\ref{eq:T00_exp}), (\ref{eq:T0i_exp}) and (\ref{eq:multipole_exp}) are not independent of each other for self-gravitating objects and require careful handling~\cite{Maggiore_vol_1}.

\section{Corrections to the energy-momentum and trajectory}
\label{sec:III}
As a first step towards computing the GW power at the 1PN order, we first calculate the correction to the energy-momentum in Eq.~(\ref{eq:T00_exp}). The energy-momentum of a perfect fluid of $N$ particles in a curved spacetime is given by~\cite{Maggiore_vol_1}
\begin{equation}\label{eq:energy_curved}
    T^{\mu\nu}(t,\textbf{x}) = \frac{1}{\sqrt{-g}}\sum_{a=1}^{N} m_{a}\frac{\mathrm{d}\tau_{a}}{\mathrm{d}t}\frac{\mathrm{d}x_{a}^{\mu}}{\mathrm{d}\tau_{a}}\frac{\mathrm{d}x_{a}^{\nu}}{\mathrm{d}\tau_{a}}\delta^{(3)}(\vec{x}-\vec{x}_{a}(t)),
\end{equation}
where $\vec{x}_{a}(t)$ is the trajectory of particle labeled by $a$, $m_{a}$ is its mass, and $\tau_{a}$ is its associated proper time. By matching the order of the metric and energy-momentum in the Einstein field equations, we see that we can use (at the order of our interest)
the standard weak field metric 
\begin{equation}\label{eq:weak_field}
    c^{2}\mathrm{d}\tau^{2} \simeq (1+2\Phi)c^{2}\mathrm{d}t^{2}- (1-2\Phi)\mathrm{d}x_{i}\mathrm{d}x^{i},
\end{equation}
where $\Phi\ll 1$ is a small perturbation to the Minkowski metric given by 
\begin{equation}
     \Phi(t,\vec{x}) = -\frac{G}{c^4}\int \mathrm{d}^{3}x'\frac{{}^{(0)}T^{00}(t,\vec{x}')}{|\vec{x}-\vec{x}'|},
\end{equation}
which is related to the Newtonian potential, $U$, via $U = -c^{2}\Phi = G M/r$, where $M$ is the total mass and $r$ is the distance between the masses. Using the metric given by Eq.~(\ref{eq:weak_field}), we find the relevant factors in Eq.~(\ref{eq:energy_curved}) to first order to give the required energy-momentum terms of Eq.~(\ref{eq:T00_exp}) as
\begin{align}
{}^{(0)}T^{00}(t,\vec{x}) &= \sum_{a=1}^{N} m_{a} c^{2}\delta^{(3)}(\vec{x}-\vec{x}_{a}(t)), \label{eq:energy_zero}\\
{}^{(2)}T^{00}(t,\vec{x}) &= \sum_{a=1}^{N} m_{a} \left(\frac{1}{2}v_{a}^{2}+\Phi c^{2}\right)\delta^{(3)}(\vec{x}-\vec{x}_{a}(t)),~~~\label{eq:energy_two}
\end{align}
where $v_{a}$ is the velocity of the particle and we can identify Eq.~(\ref{eq:energy_zero}) as the mass-energy of the particle. A similar procedure can be used to find the terms in Eq.~(\ref{eq:T0i_exp}) and in the expansion of $T^{ij}$ which is also needed to find the PN trajectory. We find 
\begin{align}
{}^{(1)}T^{0i}(t,\vec{x}) &= \sum_{a=1}^{N} m_{a}\,c \,v_{a}^{i}\,\delta^{(3)}(\vec{x}-\vec{x}_{a}(t)), \label{eq:t0i}\\
{}^{(2)}T^{ij}(t,\vec{x}) &= \sum_{a=1}^{N} m_{a}\,v_{a}^{i}\, v_{a}^{j}\,\delta^{(3)}(\vec{x}-\vec{x}_{a}(t)). \label{eq:tij}
\end{align}
The next important step in the calculation involves determining the trajectory accurate to the 1PN order, as discussed in Sec.~\ref{sec:II}. The scenario we are considering is that of a hyperbolic encounter (technically unbound scattering which reduces to hyperbola) of two compact objects (BHs or NSs) with point masses $m_1$ and $m_2$ (so $N=2$ in what follows). 

As we are interested in describing the orbit to 1PN order, we expand the trajectory and the derivative of the angle into the 0PN and 1PN contributions as 
\begin{align}
r(\phi)&= {}^{(0)}r(\phi)+{}^{(2)}r(\phi)+\mathcal{O}\left(v^4/c^4\right),\\
\dot{\phi}&= {}^{(0)}\dot{\phi}+{}^{(2)}\dot{\phi}+\mathcal{O}\left(v^4/c^4\right).
\end{align}
Then, the orbit is described at the lowest order by a Newtonian trajectory in polar coordinates as 
\begin{equation}
\label{eq:hyperbolic_traj}
{}^{(0)}r(\phi)=\frac{a(e^{2}-1)}{1+e\cos\phi},
\end{equation}
where $a$ and $b$ are the semi-major axis and impact parameter, respectively. The eccentricity is given by
\begin{equation}
 e = \sqrt{1+\frac{b^2}{a^2}} = \sqrt{1+\frac{b^2 v_0^4}{G^2 M^2}} > 1,
\end{equation}
where the total mass of the system is denoted by ${M=m_1+m_2}$, the reduced mass in a two body system is given by ${\mu = m_{1}m_{2}/M}$, and $v_0$ is the asymptotic velocity of one mass with the origin placed on the other mass.
\begin{figure}[!t]
 \centering
 \includegraphics[width = 0.8\textwidth]{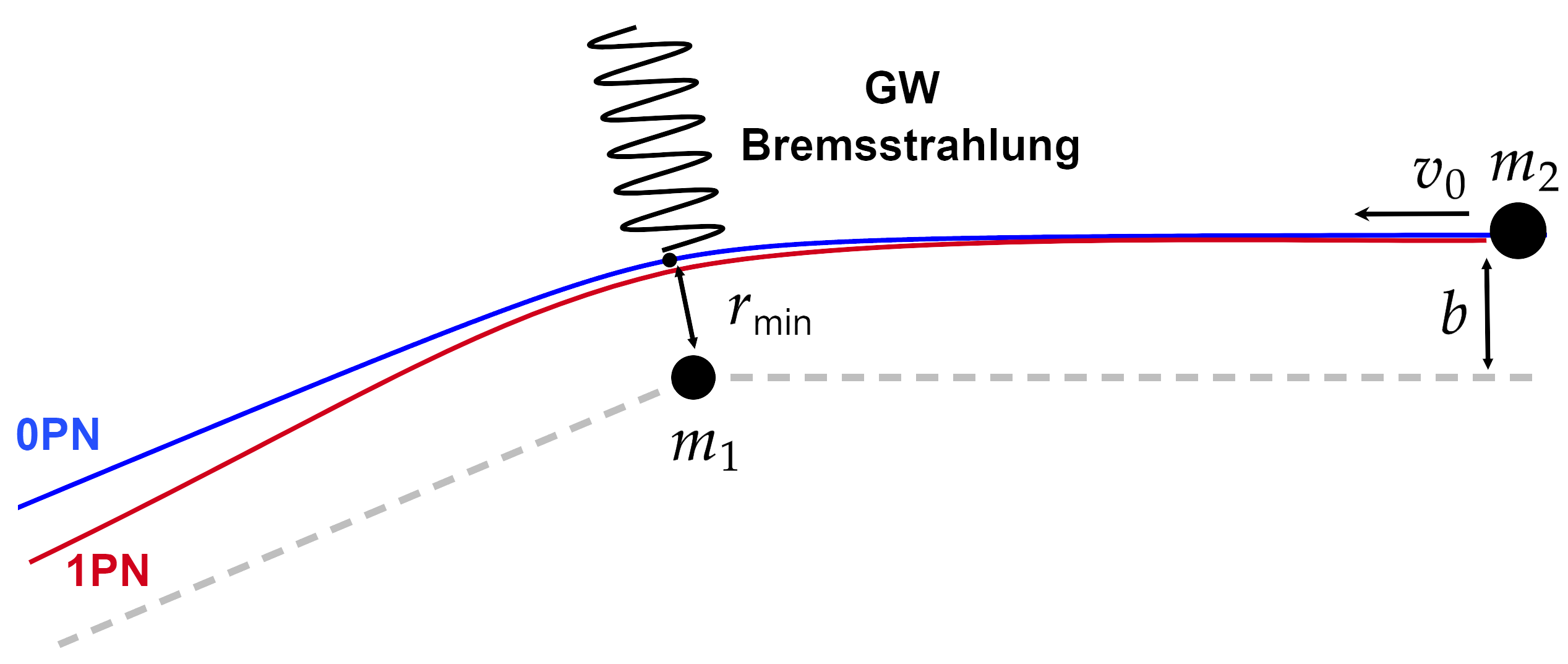}
 \caption{\justifying Hyperbolic orbit of two compact objects, where $r_{\text{min}}$ is the distance to the periapsis and $b$ is the impact parameter. The $0$PN trajectory is displayed in blue and the $1$PN in red. In Eq.~(\ref{eq:hyperbolic_traj}) the origin is located at $m_{1}$ and the $x$ axis is in the direction of the periapsis.}
 \vspace*{-1mm}
 \label{fig:hyperbolic_orbit}
\end{figure}
With reference to Fig. \ref{fig:hyperbolic_orbit}, we can define $r_{\text{min}}$ as the minimum distance of approach between the two bodies (periapsis distance), given by
\begin{equation}
 r_{\text{min}}=a(e-1).
\end{equation}
Accordingly, we can rewrite the eccentricity $e$ as 
\begin{equation}
 e = \frac{v_{0}^2 }{GM}r_{\text{min}} + 1,
\end{equation}
and it is therefore clear that all of the orbital parameters at 0PN order can be derived from the three directly measurable parameters, $M$, $v_0$ and $r_{\text{min}}$.\footnote{In what follows we will characterize the higher PN corrections using the 0PN quantities, like $e$, $a$ etc.} Also, the time derivative of the angle $\phi$, which can be derived from the angular momentum conservation, is given to 0PN via
\begin{equation}\label{eq:phi_dot}
{}^{(0)}\dot{\phi}=\frac{J}{\mu\,{}^{(0)}r^{2}(\phi)} = \frac{\sqrt{G M r_{\text{min}}(1+e)}}{{}^{(0)}r^2(\phi)},
\end{equation}
where $J/\mu= b\,v_{0}= \sqrt{G M r_{\text{min}}(1+e)}$ is the angular momentum and Eq.~(\ref{eq:phi_dot}) is the equation for the conservation of angular momentum. Similarly, the 0PN energy of the system at 0PN is
\begin{equation}
E=\frac{G\, M\,\mu}{2\,r_{\text{min}}}\,(e+1).
\end{equation}
For the 1PN terms we can use Eqs.~(\ref{eq:energy_zero}) and (\ref{eq:energy_two}), as well as the other relevant terms of the expansions of $T^{0i}$ and $T^{ij}$, i.e. Eqs.~(\ref{eq:t0i}) and (\ref{eq:tij})), to solve for the PN metric at the 1PN order and use this to find the equations of motion. The result of this procedure is 
\begin{align}
\left(\frac{\mathrm{d}(1/s)}{\mathrm{d}t}\right)^{2} &= a_{0}+a_{1}s + a_{2}s^{2}+a_{3}s^{3}, \label{eq:radius_diff}\\
\dot{\phi} &= d_{0}s^2 + d_{1}s^{3},\label{eq:phi_diff}
\end{align}
where $s\equiv G M/r$ and $a_{0}$, $a_{1}$, $a_{2}$, $a_{3}$, $d_{0}$, and $d_{1}$ are a set of coefficients which depend on the orbital energy and angular momentum of the system (see Ref.~\cite{Cho_2018} for more details) as follows
\begin{align}
a_{0} &= \frac{2\Tilde{E}}{\mu}+\frac{1}{c^2}\left(\frac{(2\Tilde{E})^2}{\mu^2}\frac{9\eta-3}{4}\right), \\
a_{1}&=2-\frac{1}{c^2}\left(\frac{2\Tilde{E}}{\mu}(6-7\eta)\right),\\
a_{2} &= -\frac{\Tilde{J}^{2}}{G^{2}M^{2} \mu^2}-\frac{1}{c^{2}}\left(\frac{2\Tilde{E}\Tilde{J}^{2}}{G^{2}M^{2} \mu^3}(3\eta-1) + 5\eta-10\right),
\end{align}
\begin{align}
a_{3} &= \frac{1}{c^2}\frac{\Tilde{J}^{2}}{G^{2}M^{2} \mu^2}(8-3\eta),\\
d_{0} &= \frac{\Tilde{J}}{G M \mu}+\frac{1}{c^2}\left(\frac{2\Tilde{E}\Tilde{J}}{G M \mu^2}\frac{3\eta-1}{2}\right),\\
d_{1} &= \frac{1}{c^{2}}\frac{\Tilde{J}}{G M \mu}(2\eta-4),
\end{align}
where $\Tilde{E}\equiv E/\mu$, $\Tilde{J}\equiv J/\mu$, and $\eta\equiv \mu /M$ is the symmetric mass ratio. Notice that the limit $c\rightarrow\infty$ returns the $0$PN expressions. Solving the system of differential equations given by Eqs.~(\ref{eq:radius_diff}) and (\ref{eq:phi_diff}), and expanding to first order in $v^2/c^2$, we find
\begin{equation}
\begin{split}
{}^{(2)}r(\phi) &= \frac{1}{c^{2}}\frac{G M }{8 e ( 1 + e \cos \phi)^2}  \biggl(2e^3 \eta  \cos 2 \phi +2e^3 (3 \eta -8)\\
&+8 e^{2} (5 \eta -13) \phi  \sin \phi +80e \eta-192e\\
&+\left(e^4 (3 \eta -1)+2 e^2 (23 \eta -59)+39 \eta -89\right) \cos \phi \biggl),
\end{split}
\end{equation}
and for the derivative of the angle
\begin{equation}\label{eq:phi_dot_1PN}
\begin{split}
{}^{(2)}\dot{\phi} &= \frac{1}{c^2} \frac{(G\,M)^{3/2} (e \cos (\phi )+1)}{4 e\,(e+1)^{5/2}\,r_\mathrm{min}^{5/2}} \biggl(\left(e^2+1\right) \biggl(3e^{2} \eta -e^{2} \\
   &-39 \eta +89\biggl) \cos \phi
   +2 e \biggl(e^2 (\eta -4) \cos 2 \phi \\
   &+e^2 (2 \eta +3)+4 e (13-5 \eta ) \phi  \sin \phi -39 \eta +89\biggl)\biggl).
   \end{split}
\end{equation}

\section{Mass quadrupole contribution}
\label{sec:IV}
In the following, we discuss the leading order term of the multipole expansion, the mass quadrupole moment, and describe how we find its $1$PN order expressions.
We then find its contribution to GW waveforms. We choose a coordinate system where the trajectory is given as
\begin{equation}
 \vec{r} = r(\phi)\left(\cos\phi, \cos\iota\sin\phi, \sin\iota\sin\phi\right).
\end{equation}
This corresponds to a plane of interaction which is the $xy$ plane rotated by $\iota$ around the $x$-axis. Moreover, the observer is selected to be in the $z$ direction. 

As explained in Sec.~(\ref{sec:II}), we include the $1$PN correction to $T^{00}$ and the trajectory when calculating the mass quadrupole moment. We therefore begin by substituting the sum of Eqs.~(\ref{eq:energy_zero}) and (\ref{eq:energy_two}) into Eq.~(\ref{eq:first_mass}) (as well as specifying that $N=2$) yielding
\begin{equation}
M^{ij} = \sum_{a=1}^{2}m_{a}\left(1+\Phi + \frac{v_{a}^{2}}{2c^{2}}\right)x_{a}^{i}x_{a}^{j},
\end{equation}
where we have integrated over the Dirac delta function. We then choose to work in the centre of mass frame defined by $m_{1}\,\vec{x}_{1}+m_{2}\,\vec{x}_{2}=\vec{0}$. The velocities of the masses can be written in an expansion of the centre of mass velocity $v$, using the metric of Eq.~(\ref{eq:weak_field}), to give 
\begin{align}
\frac{v_{1}}{c} &= \frac{m_{2}}{M}\left(\frac{v}{c}\right) + \frac{m_{2}}{M}\frac{M^{2}-m_{2}^{2}}{M^{2}}\left(\frac{v}{c}\right)^{3} + \mathcal{O}\left(v^4/c^4\right),\\
\frac{v_{2}}{c} &= -\frac{m_{1}}{M}\left(\frac{v}{c}\right) - \frac{m_{1}}{M}\frac{M^{2}-m_{1}^{2}}{M^{2}}\left(\frac{v}{c}\right)^{3} + \mathcal{O}\left(v^4/c^4\right). \ \ \ \ \ 
\end{align}
Using these results and expanding to $1$PN order we find 
\begin{equation}
M^{ij} = \mu\, r^{i}\,r^{j}\left[1+\Phi(r) +\left(\frac{M-3\mu}{2M}\right)\left(\frac{v}{c}\right)^2+\mathcal{O}(v^4/c^4)\right].
\end{equation}
Again, we decompose the quadrupole tensor and the strain to 0PN and 1PN terms as:
\begin{align}
Q^{ij}&= {}^{(0)}Q^{ij}+{}^{(2)}Q^{ij}+ \mathcal{O}\left(v^4/c^4\right),\label{eq:quad_expan}\\
h_{+,\times}&={}^{(0)}h_{+,\times}+{}^{(2)}h_{+,\times}+ \mathcal{O}\left(v^4/c^4\right).
\end{align}
The $0$PN reduced quadrupole moment in Eq.~(\ref{eq:quad_expan}) is then found to be
 \begin{equation}\label{eq:reduced_che}
 \resizebox{0.99\textwidth}{!}{
 $ {}^{(0)}Q^{kl} = \mu\,r^{2}(\phi)
 \begin{pmatrix}
 \frac{1}{6}\left(1+3\cos{2\phi}\right) & \cos\iota\cos\phi\sin\phi & \sin\iota\cos\phi\sin\phi\\
 \cos\iota\cos\phi\sin\phi & \frac{1}{12}\left(-1+3\cos2\iota-6\cos^2\iota\cos2\phi\right) & \cos\iota\sin\iota\sin^2\phi\\
 \sin\iota\cos\phi\sin\phi & \cos\iota\sin\iota\sin^2\phi & -\frac{1}{12}\left(1+3\cos2\iota+6\cos2\phi\sin^2\iota\right)
 \end{pmatrix}$},
\end{equation}
where ${}^{(0)}Q^{kl}\equiv{}^{(0)}M^{kl} -(1/3)\delta^{kl}\,{}^{(0)}M^{ii}$ and ${{}^{(0)}M^{kl}\equiv \mu \,r^{k}r^{l}}$ is understood to be the lowest order approximation to the mass moment (with $r(\phi)$ to $0$PN). Using this result, we proceed to calculate the quadrupole contribution to the GW strain in the TT gauge, given by Eq.~(\ref{eq:new_h_quad}). We find that the two polarisations of the mass quadrupole contribution for the system are given by
\begin{align}
 \begin{split}
 {}^{(0)}(h_{+})_{\text{quad}} &= -\frac{G \mu }{8 c^2 D \left(e+1\right)R} \biggl[8 e^2 \cos ^2\iota +e \left(7 \cos 2 \iota +13\right) \cos \phi\\
 &+(\cos 2 \iota +3) (e \cos 3
 \phi 
 +4 \cos 2 \phi)\biggl],
 \end{split}\\
 \begin{split}
 {}^{(0)}(h_{\times})_{\text{quad}} &= -\frac{ G \mu }{c^2 D \left(e+1\right)R} \cos \iota \sin \phi \biggl[ e \cos 2 \phi +3e+4 \cos \phi\biggl],
 \end{split}
\end{align}
where $R\equiv r_{\text{min}}/r_{s}$ and $r_{s}\equiv 2GM/c^2$. Repeating the analysis to 1PN, we find the corresponding quadrupole tensor and therefore the strain to be
\begin{align}
\begin{split}
{}^{(2)}(h_{+})_{\text{quad}} &=
 -\frac{G \mu}{8 c^2 \text{D} (e+1) R}
\biggl[8 e^2 \cos ^2\iota+e (7 \cos2\iota+13) \cos\phi \\
&+(\cos2\iota+3) (e \cos 3 \phi+4 \cos 2\phi)\biggl] \\
&+\frac{G\mu}{{128 c^2 D e (e+1)^2 R^2}}
\biggl[6 e^4 \eta  (\cos2\iota+3) \cos5\phi \\
&+4 e^3 (17 \eta -14) (\cos2\iota+3) \cos4\phi+16 e \cos2\phi  \\
&\times\Big(\left(e^2 (10 \eta -13)+41 \eta -81\right) \cos 2 \iota  
+e^2 (32 \eta -63)+3 (41 \eta -81)\Big) \\
&+\cos\phi \Big\{\Big(e^2 \big(e^2  \times(133-55 \eta ) +690 \eta -1238\big)
+7 (39 \eta -89)\Big)  \cos2\iota \\
&+e^2 \big(e^2 (55-61 \eta )
+1766 \eta -3378\big) +13 (39 \eta -89)\Big\} \\
&+\cos3\phi \Big(\big(e^2 \big(e^2 (9\eta +19) +318 \eta -506\big)
+39 \eta -89\big)\cos2\iota  \\
&+e^2 \big(e^2 \times(27 \eta -7)+954 \eta-1518\big) 
+3 (39 \eta -89)\Big) -8 e^2 (5 \eta -13) \phi \\
&\times\Big((3 \cos2\iota+17) \sin\phi -(\cos2\iota+3) \sin3\phi\Big)  \\
&+4 e \Big(e^2 \left(e^2 (2-6 \eta )+89 \eta -126\right) \\
&+\big(e^4 (2-6 \eta )+e^2 (27 \eta-10)+78 \eta -178\big) 
\cos2\iota+78 \eta -178\Big)\biggl],
\end{split}\\
\begin{split}
{}^{(2)}(h_{\times})_{\text{quad}} &= -\frac{ G \mu }{c^2 D \left(e+1\right)R} \cos \iota \sin \phi \biggl[ e (\cos2\phi+3)
+4 \cos \phi\biggl]  \\
&+\frac{ G \mu }{32c^2 D \left(e+1\right)^2R^2}
\cos\iota \biggl[6 e^4 \eta  \sin5\phi
+4 e^3 (17 \eta -14)\sin4\phi \\
&+8 e \left(e^2 (21 \eta -38)+82 \eta -162\right) \sin2\phi \\
&+8 e^2 (5 \eta -13) \phi  (4 e+5 \cos\phi-\cos3\phi) \\
&-\big(e^4 (17\eta +49)+e^2 (1154-614 \eta )
-195 \eta +445\big)\sin\phi \\
&+\big(e^4 (9 \eta +3)+e^2 (318 \eta -506)
+39 \eta-89\big) \sin3\phi \biggl].
\end{split}
\end{align}

\section{Mass octupole and current quadrupole \label{sec:V}}
We construct the mass octupole of Eq.~(\ref{eq:mass_octupole}) using the mass moments given by $M^{klm}=\mu~r^{k}r^{l}r^{m}$. Since the system is observed in the $z$ direction, the contraction with $n_{m}$ in Eq.~(\ref{eq:mass_octupole_contribution}) means that we only need to consider the $\mathcal{O}^{kl3}$ components which we list here for ease of reference
\begin{align}\label{eq:new_mass_octupole}
    \mathcal{O}^{113} &= \sqrt{1-4\eta}\mu r^{3} \frac{1}{20}\sin \iota (\sin \phi +5 \sin 3 \phi ),\\
    \mathcal{O}^{123} &= \sqrt{1-4\eta}\mu r^{3}\sin \iota \cos \iota \sin ^2\phi \cos \phi ,\\
    \mathcal{O}^{133} &= -\sqrt{1-4\eta}\mu r^{3}\frac{1}{40}\left(\cos \phi \left(20 \cos 2 \iota \sin ^2\phi +3\right)+5 \cos 3 \phi \right),\\
    \mathcal{O}^{213} &= \sqrt{1-4\eta}\mu r^{3}\sin \iota \cos \iota \sin ^2\phi \cos \phi,\\
    \mathcal{O}^{223} &= \sqrt{1-4\eta}\mu r^{3}\frac{1}{20}\sin \iota \sin \phi \left(5 \cos 2 \iota -10 \cos ^2\iota \cos 2 \phi +1\right),\\
    \mathcal{O}^{233} &= -\sqrt{1-4\eta}\mu r^{3}\frac{1}{80} \left(20 \sin ^2\iota \cos \iota \sin 3 \phi +(\cos \iota +15 \cos 3 \iota ) \sin \phi\right),\\
    \mathcal{O}^{313} &= -\sqrt{1-4\eta}\mu r^{3}\frac{1}{40}\left(\cos \phi \left(20 \cos 2 \iota \sin ^2\phi +3\right)+5 \cos 3 \phi \right),\\
    \mathcal{O}^{323} &=-\sqrt{1-4\eta}\mu r^{3}\frac{1}{80} \left(20 \sin ^2\iota \cos \iota \sin 3 \phi +(\cos \iota +15 \cos 3 \iota ) \sin \phi\right),\\
    \mathcal{O}^{333} &= -\sqrt{1-4\eta}\mu r^{3}\frac{1}{80}\left(20 \sin ^3\iota \sin 3 \phi +3 (\sin \iota +5 \sin 3 \iota ) \sin \phi \right).\label{eq:new_mass_octupole_2}
\end{align}
We can therefore compute the mass octupole contribution to the GW strain in the TT gauge, see Eq.~\eqref{eq:mass_octupole_contribution}, by taking the third time derivative of Eqs.~(\ref{eq:new_mass_octupole})-(\ref{eq:new_mass_octupole_2}). We find that the two
polarisations of the mass octupole contribution are
\begin{align}
\begin{split}
(h_{+})_{\text{oct}} &=\frac{G \mu \sqrt{1-4\eta} \sin \iota }{192\sqrt{2} c^2 D \left(e+1\right)^{3/2}R^{3/2}}
\biggl[4 e \Big\{\left(3-24 e^2\right) \cos 2 \iota -24 e^2+1 \Big\}\\
&-9 e^2 (\cos 2\iota +3) \cos 5 \phi 
+2 \Big\{\left(6-111 e^2\right) \cos 2\iota -177 e^2+2 \Big\} \cos \phi\\
&-\Big\{ 3 \left(19e^2+36\right) \cos 2 \iota +131 e^2+324 \Big\} \cos 3 \phi \\
&-80 e (3 \cos 2 \iota +7) \cos 2 \phi 
-60 e (\cos 2\iota +3) \cos 4 \phi \biggl],
\end{split}
\end{align}
and
\begin{align}
\begin{split}
(h_{\times})_{\text{oct}} &= -\frac{G \mu \sqrt{1-4\eta} \sin 2 \iota \sin \phi }{48\sqrt{2} c^2 D \left(e+1\right)^{3/2}R^{3/2}} \biggl[9 e^2 \cos 4 \phi \\
&+4 \left(14 e^2+27\right) \cos 2 \phi+95 e^2
+260 e \cos \phi +60 e \cos 3 \phi +52\biggl].
\end{split}
\end{align}
We now form the current quadrupole Eq.~(\ref{eq:current_quad_contribution}) for the system, which is constructed from the moment of momentum ${P^{k,lm}=\mu~\dot{r}^{k}r^{l}r^{3}}$. Similarly to the mass octupole, we only need the $\mathcal{Z}^{kl3}$ component, which is given by
\begin{equation}\label{eq:current_quadrupole}\mathcal{Z}^{kl3}= \sqrt{1-4\eta}\mu b v_{0} r(\phi)
 \begin{pmatrix}
 -2\sin\iota\cos\phi & - \cos\iota\sin\iota\sin \phi &- \sin^2\iota\sin \phi\\
 - \cos\iota\sin\iota\sin \phi & 0 & 0 \\
 - \sin^2\iota\sin \phi & 0 & 0 
\end{pmatrix},
\end{equation}
where we have made use of the formula for $\dot{\phi}$ (Eq.~(\ref{eq:phi_dot})). We can therefore calculate, by finding the second time derivative of Eq.~(\ref{eq:current_quadrupole}), that the two polarisations for the current quadrupole contribution of system are given by 
\begin{align}
 \begin{split}
 (h_{+})_{\text{cur.quad.}} = \frac{\sqrt{2-8\eta} G \mu \sin\iota}{3 c^2 D \left(e+1\right)^{3/2}R^{3/2}} \cos \phi ~ (1+e \cos \phi )^2,
 \end{split}\\
 \begin{split}
 (h_{\times})_{\text{cur.quad.}} = \frac{\sqrt{2-8\eta} G \mu \sin2\iota}{6 c^2 D \left(e+1\right)^{3/2}R^{3/2}} \sin \phi ~(1+e \cos \phi)^2.
 \end{split}
\end{align}
\section{Power emission and total energy emitted at different orders}
\label{sec:VI}
\begin{figure*}[!t]
\centering
\includegraphics[width=0.6\textwidth]{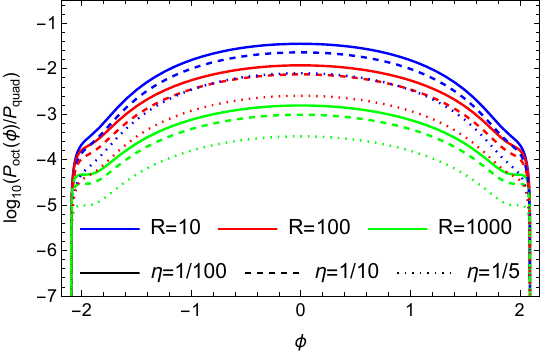}\vspace{2mm}
\includegraphics[width=0.6\textwidth]{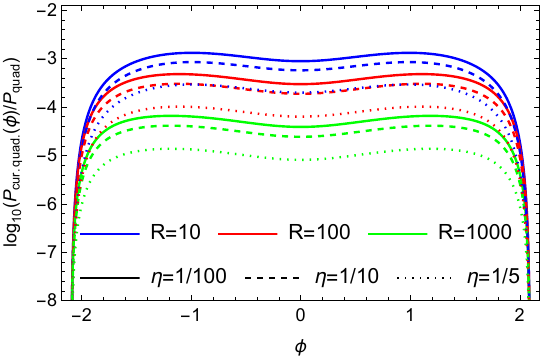}\vspace{2mm}
\includegraphics[width=0.6\textwidth]{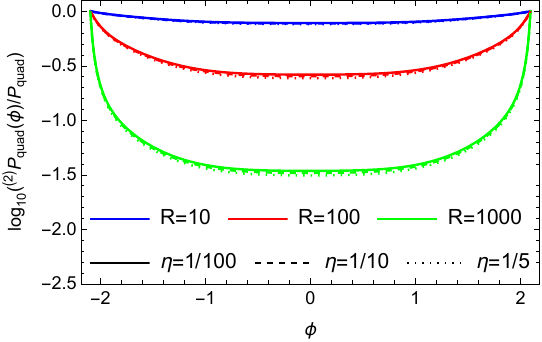}
\caption{\justifying Plots illustrating the logarithmic ratios of octupole power (top), the current quadrupole power (middle) and the 1PN contribution to the quadrupole (bottom), relative to the combined 0PN and 1PN quadrupole power. We plot them with a fixed eccentricity $e=2$ for various values of the parameter $R\in[10,10^2, 10^3]$ (the ratio of the periapsis distance $r_\mathrm{min}$ to the Schwarzschild radius $r_s$) and $\eta\in [1/100, 1/10, 1/5]$ (symmetric mass ratio) as a function of the angle of the orbit $\phi$. Note the weak dependence on $\eta$ in the plots of 1PN quadrupole.\label{fig:pows}}
\end{figure*}

By using the standard expressions for the power emission at different orders, given in Eq.~(\ref{eq:power_nextto}), we find the power as a function of angle from the quadrupole (expressed as $P_\mathrm{quad}={}^{(0)}P_\mathrm{quad}+{}^{(2)}P_\mathrm{quad}$), octupole, and current quadrupole as
\begin{align}
\begin{split}
{}^{(0)}P_\mathrm{quad} &=\frac{G\,\mu^2\, c}{30\,(1+e)^5\, R^5\,r_s^2}\biggl[1+e\,\cos\phi\biggl]^4 
 \left(11\,e^2\,\cos2\phi+13 e^2+48\,e\, \cos\phi+24\right)\\
\end{split}
\end{align}
\begin{align}
\begin{split}
{}^{(2)}P_\mathrm{quad} &= \frac{G\,\mu^2\, c}{480\,e\,(1+e)^6\, R^6\,r_s^2} \biggl[1+e\,\cos\phi\biggl]^3 \\
&\times\biggl[-16 e^2 (5 \eta - 13) \phi \sin (\phi) \big(33 e^2 \cos 2 \phi + 37 e^2 \\
&+ 142 e \cos \phi + 72\big) + \big(e^6 (181 \eta - 33)
+ 6 e^4 (5327 - 2619 \eta) \\
&+ 33 e^2 (3095 - 1481 \eta) + 144 (89 - 39 \eta)\big) \cos \phi \\ 
&+ e \Big(3 e^5 (20 - 21 \eta) \cos 5 \phi + 2 e^4 (500 - 441 \eta) \cos 4 \phi \\
&+ e \Big(e^2 \left(e^2 (117 - 70 \eta) - 6126 \eta + 9750\right)
+ 33 (89 - 39 \eta)\Big) \cos3 \phi\\
&+ 2 e^2 \left(e^2 (1405 - 526 \eta) - 11686 \eta + 24198\right) \\
&+ \big(e^4 (3774 - 2002 \eta) + e^2 (45492 - 23668 \eta) \\
&+ 142 (89 - 39 \eta)\big) \cos 2 \phi - 21054 \eta + 45634\Big) \biggl],
\end{split}\\
\begin{split}
P_\mathrm{oct} &= \frac{G\,\mu ^2\,c\, (1-4\eta)}{20160 \,(1+e)^6\, R^6\,r_s^2}\,\biggl[1+e \cos\phi\biggl]^4 \\
&\times \biggl[3101 e^4+5 e \Big\{8 \left(653 e^2+1012\right) \cos \phi
+ e \{ \left(796 e^2+5624\right) \cos2\phi \\
&+5 e (39 e \cos4\phi+344 \cos3\phi) \} \Big\} 
+28088 e^2+10936\biggl],
\end{split}\\
\begin{split}
P_\mathrm{cur.quad.} &= \frac{G \mu ^2c\, (1-4\eta)}{180 (1+e)^6\,R^6\, r_s^2}\biggl[1+e \cos\phi\biggl]^6 
\left(-3 e^2 \cos2\phi+5e^2+4 e \cos\phi+2\right).
\end{split}
\end{align}
As an example, in Fig.~\ref{fig:pows} we show the logarithmic ratios of the power of 
the octupole, current quadrupole, and 1PN
contribution to the quadrupole with respect to that of the combined 0PN and 1PN quadrupole\footnote{
We use the combined 0PN and 1PN quadrupole as the basis for comparison because it provides the most accurate prediction for the quadrupole within the PN order considered in this paper. Current matched-filtering searches typically focus only on quadrupole radiation (with higher-order PN corrections). Therefore, the relative differences presented here approximately indicate the effects of multipole corrections (mass octupole and current quadrupole) in comparison to the matched-filtering templates used in the current analysis.
}. We fix the eccentricity to be
$e=2$ and plot the ratios for various values of the parameter $R$ and $\eta$, as a function of the angle of the orbit $\phi$. 
As can be seen, the ratios depend strongly on $R$, which is the ratio of the periapsis distance $r_\mathrm{min}$ to the Schwarzschild radius $r_s$. For small values of $R$, the semi-Keplerian approximation breaks down, while for $R>10$ it seems that the octupole and current quadrupole are subdominant to the quadrupole, as expected. 
The ratios also exhibit a dependence on the symmetric mass ratio $\eta$. Specifically, the contributions from the octupole and current quadrupole become larger for small $\eta$, corresponding to a large mass asymmetry between the two objects. We see also that the octupole and current quadrupole do not contribute when $\eta\simeq 1/4$ ie $m_{1}\simeq m_{2}$. Also, comparing the 
left and middle panels, we find that the contribution of the octupole is roughly one order of magnitude larger than the current quadrupole. 
From the right panel, we see that the next-to-leading order quadrupole contribution is far larger than the octupole and quadrupole current contributions when we fix $e=2$. It is worth noting that the edges at $\phi \sim \pm 2$ correspond to the limit where the BHs are infinitely far away and the curves show non-trivial behavior as both the numerator and denominator rapidly approach zero.


\begin{figure*}[!t]
\centering
\vspace{10mm}
\includegraphics[width=0.62\textwidth]{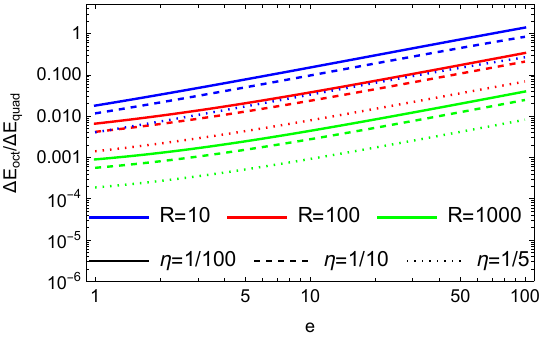}\vspace{2mm}
\includegraphics[width=0.62\textwidth]{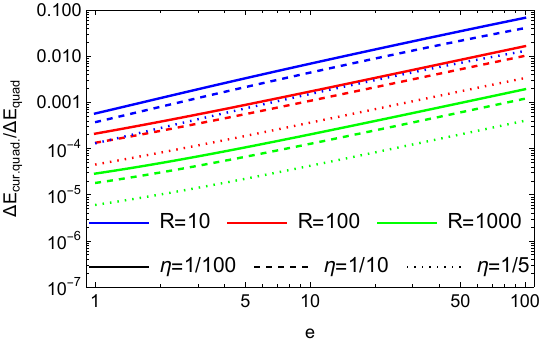}\vspace{2mm}
\includegraphics[width=0.62\textwidth]{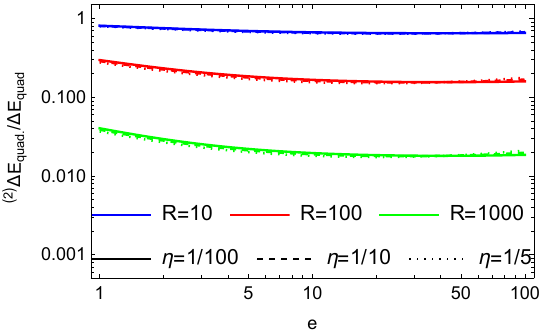}
\caption{\justifying Plots of the ratios of the total energy emitted at each order with respect to that of the quadrupole, as a function of the eccentricity $e$ and for various values of the parameters $R$ and $\eta$. The order of the plots and the range of the parameters $R$ and $\eta$ are the same as in Fig.~\ref{fig:pows}.
\label{fig:rats}}
\end{figure*}

Integrating over the orbit, i.e., over the asymptotic limits of $[-\phi_0,\phi_0]$, where $\phi_0=\arccos(-1/e)$, we can find the total energy emitted at each order as 
\begin{equation}
\Delta E \equiv \int_{-\infty}^{\infty}\, P(t)\,\mathrm{d}t \
= \int_{-\phi_0}^{\phi_0}\,\frac{P(\phi)}{\dot{\phi}}\,\mathrm{d}\phi.
\end{equation}
Then, at each order we have:
\ba
{}^{(0)}\Delta E_\mathrm{quad} &=& \frac{c^4 r_s \eta^2}{360 \sqrt{2}\,(1+e)^{7/2}\,R^{7/2}\, G}\biggl[\sqrt{e^2-1} \left(673 e^2+602\right)\nn\\
&+&3 \left(37 e^4+292 e^2+96\right) \sec^{-1}(-e)\biggl]
\ea
\ba
{}^{(2)}\Delta E_\mathrm{quad} &=& \frac{c^4 r_s \eta^2}{7200 \sqrt{2}\,(1+e)^{9/2}\,R^{9/2}\, G} \biggl[\sqrt{e^2-1} \nn\\
&\times&\Big(7 e^6 (8546-4215 \eta )+e^4 (1021501-510505 \eta )\nn\\
&+&17 e^2 (60891-28525 \eta )-360 (39 \eta -89) \Big)\nn\\
&+&15 e^2 \Big(e^6 (35 \eta +2)+e^4 (18033-9195 \eta )\nn\\
&+&e^2 (85613-41835 \eta )-18270 \eta +39586\Big)
\sec^{-1}(-e) \biggl],
\ea
where for the quadrupole we have the total energy $\Delta E_\mathrm{quad}={}^{(0)}\Delta E_\mathrm{quad}+{}^{(2)}\Delta E_\mathrm{quad}$ as the sum of the 0PN and 1PN terms. For the other orders we find
\ba
\Delta E_\mathrm{oct} &=& \frac{c^4 r_s \eta^2\,(1-4\eta)}{120960 \sqrt{2}\,(1+e)^{9/2}\,R^{9/2}\, G} \nn\\
&\times& \biggl[ \sqrt{e^2-1} \left(148439 e^4+521072 e^2+157364\right)\nn\\
&+&3 (5091 e^6+100590 e^4+148072 e^2
+21872) \sec^{-1}(-e)\biggl],
\ea
\ba
\Delta E_\mathrm{cur.quad.} &=& \frac{c^4 r_s \eta^2\,(1-4\eta)}{4320 \sqrt{2}\,(1+e)^{9/2}\,R^{9/2}\, G}  \nn\\
&\times& \biggl[\sqrt{e^2-1} \left(229 e^4+592 e^2+124\right) \nn\\
&+&3 \left(9 e^6+138 e^4+152 e^2+16\right) \sec^{-1}(-e)\biggl]. \nn\\
\ea
Finally, having defined the energy contribution at each order, we can also estimate the ratios of the energies at different orders, which after expanding around $e\simeq1$, can be found to be  
\ba 
\frac{\Delta E_\mathrm{oct}}{\Delta E_\mathrm{quad}} &\simeq&\frac{13125 (1-4 \eta)}{4 (-69265 \,\eta +3400 R+143234)}\nn \\
&+&\frac{-49589995 \eta +1131850 R+106702547}{6 (-69265 \,\eta +3400 R+143234)^2}
\times 25 (e-1) (1-4 \eta) \nn\\
&+&\mathcal{O}(e^{2}),\label{eq:DE_oct}
\ea 
\ba
\frac{\Delta E_\mathrm{cur.quad.}}{\Delta E_\mathrm{quad}} &\simeq& \frac{105-420 \eta }{-69265 \,\eta +3400 R+143234}\nn \\
&+&\frac{-359365 \,\eta +10198 R+766877}{3 (-69265 \,\eta +3400 R+143234)^2}
\times70 (e-1) (1-4 \eta) \nn \\
&+&\mathcal{O}(e^{2}),\label{eq:DE_qc}
\ea
\ba
\frac{{}^{(2)}\Delta E_\mathrm{quad.}}{\Delta E_\mathrm{quad}} &\simeq &1-\frac{3400 R}{-69265 \,\eta +3400 R+143234}\nn \\
&+&\frac{4 (e-1) (28637585 \eta -63704526) R}{(-69265\, \eta +3400 R+143234)^2} + \mathcal{O}(e^{2}).~~~~\label{eq:DE_quad2}
\ea
As can be seen, all three ratios depend only on $e$, $\eta$, and $R$.

Next, we show plots of these ratios in Fig.~\ref{fig:rats} for various values of the quantities $R$ and $\eta$ as a function of the eccentricity $e$. 
The ratios are approximately linear with respect to the eccentricity, as predicted by Eqs.~\eqref{eq:DE_oct}-\eqref{eq:DE_qc}, and they are important as they can tell us when the contributions from the octupole and current quadrupole are dominant. 
For high eccentricities (when the bodies become relativistic) and small periapsis distance, the total energy emitted from the octupole and current quadrupole orders becomes equal or higher than the quadrupole, thus signifying the collapse of the weak-field/low-velocity approximation. 

We note that the 1PN correction in the quadrupole moment $Q_{ij}$ contributes at the same order as the octupole and current quadrupole terms in Eq.~(\ref{eq:power_nextto})~\cite{Maggiore_vol_1}. 
From Fig.~\ref{fig:rats}, we find that ${}^{(2)}\Delta E_\mathrm{quad}$ is mostly larger than $\Delta E_\mathrm{oct}$ and $\Delta E_\mathrm{cur.quad}$, while the octupole and current quadrupole contributions becomes comparable to the 1PN correction in the quadrupole moment when the eccentricity is large. In the extreme cases, e.g. for $e\sim50$ and $R\sim10$, the ratio of the total energy emitted becomes order one, in which case 
our setup becomes highly relativistic and we would need to use numerical relativity for precise predictions. 

Recently, searches for hyperbolic encounters of compact objects in the third LIGO-Virgo-KAGRA observing run were performed in Ref.~\cite{Bini:2023gaj} using a 3PN-accurate quasi-Keplerian parametrization only at the quadrupolar order. To perform the model-informed search, injections were made with symmetric mass ratios of up to $\eta\sim 4/25\simeq 0.16$ and impact parameters even as low as $30\,r_s$ (or in our notation $R\sim30$), 
while the eccentricity range is relatively narrow $[1.05, 1.6]$. As seen in Figs.~\ref{fig:pows} and \ref{fig:rats}, this scenario roughly corresponds to the blue dashed curve and implies that the octupole is around $1\%$ of the combined 0PN and 1PN quadrupole energy emitted by the system. This difference might not be noticeable with current data, but will be more important in the near future in the case of next-generation GW detectors.

Finally, the power spectrum can be obtained by Fourier transforming the energy emitted in time domain by following the procedure of Ref.~\cite{Garc_a_Bellido_2018}. In a nutshell, we can use Parseval's theorem to define the power in the frequency domain as
\be\label{eq:DE}
\Delta E = \int_{-\infty}^{\infty} P(t)\,\mathrm{d}t = \frac{1}{\pi}\int_0^\infty P(\omega)\,\mathrm{d}\omega.
\ee
For example, we can see that in Fourier space the power for the quadrupole is given by
\begin{equation}
P_\mathrm{quad}(\omega) = \frac{G}{5\, c^5}~\sum_{i,j} |\widehat{\dddot{Q}_{ij}}|^2
=\frac{G}{5\,c^5}~\omega^6 \sum_{i,j} |\widehat{Q_{ij}}|^2,
\end{equation}
where $\widehat{Q_{ij}}$ is the Fourier transform of the quadrupole momentum tensor $Q_{ij}$.

Similarly, for the octupole and current quadrupole, we find:
\begin{equation}
P_\mathrm{oct}(\omega)=\frac{G}{189\, c^7}~\sum_{i,j,k} |\widehat{\ddddot{\mathcal{O}}_{ijk}}|^2
=\frac{G}{189\, c^7}~\omega^8 \sum_{i,j,k} |\widehat{\mathcal{O}_{ijk}}|^2,
\end{equation}
and 
\begin{equation}
P_\mathrm{cur.quad.}(\omega)=\frac{16\,G}{45\, c^7}~\sum_{i,j} |\widehat{\dddot{\mathcal{J}}_{ij}}|^2
=\frac{16\,G}{45\, c^7}~\omega^6 \sum_{i,j} |\widehat{\mathcal{J}_{ij}}|^2.
\end{equation}
Introducing the dimensionless variable $\nu=\kappa\,\omega$, where $\kappa = \sqrt{a^3/G M}$, we can find analytic (but quite long) expressions for the power in the frequency domain for the different terms in the expansion, in terms of Hankel functions (see Refs.~\cite{Garc_a_Bellido_2018,Grobner:2020fnb} for the expressions for the quadrupole or Ref.~\cite{Caldarola:2023ipo} for the case including orbital precession effects).

Given the complexity of the final expressions for the 1PN quadrupole and the orbit, for the quadrupole power, we opt for a brute-force numerical calculation avoiding any approximation. To do so, we numerically Fourier transform the quadrupole tensor of Eq.~\eqref{eq:quad_expan} using the 1PN expressions for the orbit.

\begin{figure}[!t]
 \centering
 \includegraphics[width = 0.7\textwidth]{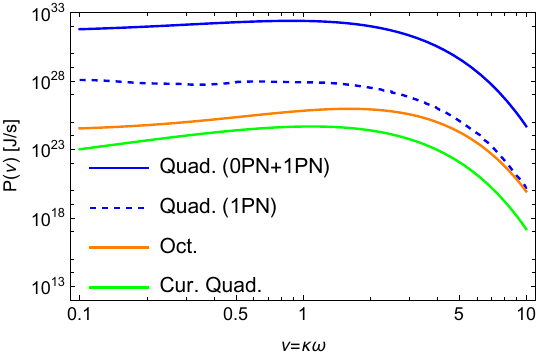}
 \caption{\justifying Plot of the emitted power in frequency domain for the different contributions, namely the quadrupole (blue line), octupole (orange line) and current quadrupole (green line), in terms of the dimensionless frequency $\nu=\kappa\,\omega$, where $\kappa = \sqrt{a^3/G M}$. The orbital parameters are set to be $m_1=10M_\odot$, $m_2=30M_\odot$, $v_0=10^{-3}c$, $b=1$AU, which correspond to $e\approx 2.72$ and $\eta\approx 0.19$.}
 \vspace*{-1mm} \label{fig:power_comparison}
\end{figure}

In Fig.~\ref{fig:power_comparison}, we show the power spectra of the different contributions as a function of the dimensionless frequency $\nu$, for some characteristic values of the parameters, namely $m_1=10M_\odot$, $m_2=30M_\odot$, $v_0=10^{-3}c$, $b=1$AU, which correspond to $e\approx 2.72$ and $\eta \approx 0.19$.  As expected, the mass quadrupole moment is the dominant contribution, while the higher-order moments, such as the mass octupole and the current quadrupole moments, can contribute but they are typically weaker by several orders of magnitude. Also, we note that the peaks of the spectra for octupole and current quadrupole are shifted to higher frequencies with respect to the quadrupole.

\section{Conclusions}
\label{sec:Conclusions}
Among the exciting new research made possible by the detection of GWs, GW bursts from close hyperbolic encounters (CHEs), in particular, have the possibility to lead to new insights into the dynamics of tightly bound BH clusters. The detection of burst events could also be used to probe the BH populations. It is, therefore, important that the GW waveforms are characterised, since these are likely to be detected in future generations of GW detectors and are far less well-studied when compared to GWs produced by merger events.

In this work, we studied GWs from CHEs and derived both the mass octupole and current quadrupole radiation. Specifically, the calculation of GW power and total energy emitted represents a novel result crucial for understanding the features of waveforms likely to be observed in the near future.

We began by revisiting the calculation of the mass quadrupole, extending it to the next-to-leading order, which aligns with the order of the mass octupole and current quadrupole contributions. Subsequently, we delved into the radiation from the mass octupole and current quadrupole. This involved deriving expressions for the emitted power and integrating over the orbit to determine the total energy emitted in both the time and frequency domains. We then compared these results with those from the mass quadrupole contribution by examining their ratio.

Finally, we identified specific configurations of hyperbolic systems characterized by low values of $R$, low $\eta$, and high $e$, where the next-to-leading order correction could be significant in searches for CHE events. This underscores the importance of accounting for such corrections in the analysis of burst events, especially as we anticipate detecting such events with high waveform accuracy using next-generation GW detectors.

\section*{Acknowledgements}
The authors would like to thank Michele Maggiore, Juan Garc\'ia-Bellido, and Gonzalo Morr\'as for helpful discussions and acknowledge support from the research project PID2021-123012NB-C43 and the Spanish Research Agency (Agencia
Estatal de Investigaci\'on) through the Grant IFT Centro de Excelencia Severo Ochoa No CEX2020-001007-S, funded by MCIN/AEI/10.13039/501100011033. M.C. acknowledges support from a SO PhD fellowship. S.K. is supported by the Spanish Atracci\'on de Talento contract no. 2019-T1/TIC-13177 granted by Comunidad de Madrid, the I+D grant PID2020-118159GA-C42 funded by MCIN/AEI/10.13039/501100011033, the Consolidaci\'on Investigadora 2022 grant CNS2022-135211 and the i-LINK 2021 grant LINKA20416 of CSIC, and Japan Society for the Promotion of Science (JSPS) KAKENHI Grant no. JP20H01899, JP20H05853, JP23H00110, and JP24K00624.

\section*{References}
\bibliographystyle{iopart-num}
\bibliography{HEoctupole}

\end{document}